\documentstyle[12pt,amsfonts]{article}
\def\one{1\hskip-.37em 1}
\def\ir{{\rm I}\hskip-.2em{\rm R}}
\def\half{\textstyle{\frac{1}{2}}}

\def\H{{\cal H}}
\def\D{{\cal D}}
\def\tr{{\rm tr}}
\def\E{{\rm I}\hskip-.2em{\rm E}}
\def\ra{\rightarrow}
\def\tint{{\textstyle\int}}
\def\d{\partial}
\def\o{\overline}
\def\a{\alpha}
\def\b{\begin{eqnarray*}}     
\def\e{\end{eqnarray*}}       
\def\bn{\begin{eqnarray}}     
\def\en{\end{eqnarray}}       
\def\<{\langle}
\def\>{\rangle}

\def\no{\nonumber}

\def\{{\lbrace}
\def\}{\rbrace}
\begin{document}

\title{Noncanonical Quantization of Gravity. I.\\
   Foundations of Affine Quantum Gravity}
\author{John R. Klauder\\
Departments of Physics and Mathematics\\
University of Florida\\
Gainesville, Fl  32611}
\date{}
\maketitle
\begin{abstract}
The nature of the classical canonical phase-space variables for gravity 
suggests that the associated quantum field operators should obey affine 
commutation relations rather than canonical commutation relations. Prior 
to the introduction of constraints, a primary kinematical representation 
is derived in the form of a reproducing kernel and its associated 
reproducing kernel Hilbert space. Constraints are introduced following the 
projection operator method which involves no gauge fixing, no complicated 
moduli space, nor any auxiliary fields. The result, which is only 
qualitatively sketched in the present paper, involves another reproducing 
kernel with which inner products are defined for the physical Hilbert 
space and which is obtained through a reduction of the original reproducing 
kernel.  Several of the steps involved in this general analysis are 
illustrated by means of analogous steps applied to one-dimensional 
quantum mechanical models. These toy models help in motivating and 
understanding the analysis in the case of gravity.
\end{abstract}
\section{Introduction}
General relativity is, in certain ways, fundamentally different than 
most other physically relevant classical field theories, and the same 
remark applies to attempts to provide associated quantum formulations. 
The space-time metric $g_{\mu\nu}(x)$, $x\in\ir^4$, $\,\mu,\nu=0,1,2,3$, 
possesses a signature requirement that is incompatible with the space of 
metrics being a linear vector space.\footnote{Although, we assume a $3+1$ 
theory of gravity for illustrative purposes, it is straightforward to 
generalize to an $s+1$ theory as well, $s\ge1$.} The inverse metric 
$g^{\sigma\mu}(x)$, defined so that $g^{\sigma\mu}(x)g_{\mu\nu}(x)=
\delta^\sigma_\nu$, is classically trivial but it is quantum mechanically 
challenged since the left-hand side involves the product of two 
operator-valued distributions. Moreover, the spatial and temporal 
constraints that hold at each space-time point classically close 
algebraically, but they exhibit an anomaly when quantized. In effect, 
this fact changes the constraints from first class (classically) to 
second class (quantum mechanically). And, of course, there is the 
well-known fact that unlike other theories which take place on a fixed 
space-time stage, the theory of gravity involves the dynamics of the 
space-time stage itself. The purpose of this article is to discuss some 
basic issues surrounding quantum gravity from a viewpoint different than  
traditional ones.\footnote{The closest work in spirit to that discussed 
in this paper is that of the author \cite{kla}, Isham and Kakas \cite{ish}, 
and especially Pilati \cite{pli}.  See Sec.~5 for an extensive discussion. 
We do not directly comment on current schemes for quantizing gravity.} 

Let us outline the general approach we shall adopt. First, we focus on 
basic kinematics and the quantum theory of positive definite, $3\times3$ 
matrix-valued field variables and associated noncanonical ``conjugate'' 
field variables, designed to offer an initial class of coherent states and 
coherent-state induced Hilbert space representations. In this step it is 
noteworthy that, besides the signature issue for the metric, the existence 
of an operator  $g^{jk}(x)$, $j,k=1,2,3$, inverse to $g_{kl}(x)$ is shown, 
such that, when suitably defined, the equation $g^{jk}(x)g_{kl}(x)=
\delta^j_k$ is fulfilled. Second, we introduce the spatial and temporal 
constraints in the projection operator approach  recently developed by the 
author and others \cite{kla1,sha, gov,kla2}. This procedure has the 
advantage of working entirely with the classical degrees of freedom, 
including the $c$-number Lagrange multipliers---specifically, the lapse 
and shift functions. It is not necessary to introduce additional fields 
(e.g., ghosts with false statistics), nor choose gauges, nor pass to moduli 
spaces, etc. Initially, the constraints are imposed in a regularized fashion. 
Subsequently, the removal of the regularization is analyzed, a process which 
often involves an automatic change of Hilbert-space representation. Assuming 
that the limit removing the regularization exists, the physical Hilbert 
space that arises is then, generally speaking, best described as a 
reproducing kernel Hilbert space that emerges from the reduction of the 
original 
reproducing kernel. 

Before we undertake any discussion of gravity, however, we sketch in Sec.~2 
the key concepts as applied to some simple, few degree-of-freedom systems. 
In Sec.~3 we construct a suitable kinematical framework for quantum gravity, 
while in Sec.~4 we analyze the introduction of constraints.  Section 5 
contains a general discussion about operator representations and constraints 
in relation to reparameterization invariance.
 In Part II of this work, the analysis of gravitational constraints 
is discussed in detail. In addition, the classical limit of the affine 
gravitational quantum theory developed in Secs.~3-4 will be discussed 
and compared with classical gravity.

\section{Elementary Illustration of Key Concepts}
As is generally well known, there is only one irreducible representation up 
to unitary equivalence of canonical, self-adjoint operators $P$ and $Q$ 
satisfying the Weyl form of the canonical commutation relations. This 
representation, equivalent to the Schr\"odinger representation, implies 
that  the spectrum of both $P$ and $Q$ cover the whole real line. Such 
operator degrees of freedom are appropriate for many systems with a finite 
or an infinite number of degrees of freedom, but they are inappropriate for 
gravity. The reason for this is that the classical $3\times3$ metric is 
strictly positive definite and the associated quantum field operator cannot 
be represented by an operator whose spectrum is unbounded above and below. 
Instead, of the usual relation $[Q,P]=i$, with $\hbar=1$, one is led to 
consider an {\it affine commutation relation} \cite{kla65}, which for a 
single degree of freedom takes the form 
  \bn [Q,D]=iQ\;. \en
Here $D\equiv(PQ+QP)/2$ denotes the dilation operator, and it follows 
\cite{gel,ask} that solutions of the affine commutation relations exist 
with irreducible, self-adjoint operators $D$ and $Q$ for which---and this 
is the important part---$Q>0$. $(\!\!($There are two other inequivalent 
self-adjoint solutions, one where $Q<0$, which is rather like the 
representation of interest, and another for which $Q=0$ \cite{gel}. Neither 
of these representations will be of interest in this article.$)\!\!)$ Even 
though the operator $P$ is only a symmetric operator which has no 
self-adjoint extension, the introduction of the self-adjoint operator 
$D$ provides the substitute commutation relation given above. These two 
commutation relations are not in conflict since the affine commutation 
relation follows directly from the Heisenberg commutation relation simply 
by multiplication of the latter by $Q$.  We note that an analog of the 
affine variables will be used in the case of the gravitational field to 
maintain the positivity of the local quantum field operator for the 
$3\times3$ spatial metric.

Continuing with the one-dimensional example, and based on self-adjoint 
operators that satisfy the affine commutation relation, let us introduce 
{\it affine coherent states} \cite{ask2}, $|p,q\>\in{\frak H}$, defined 
by the expression
  \bn |p,q\>\equiv e^{ipQ}e^{-i\ln(q)D}\,|\eta\>\;,\hskip1cm-\infty<p<
\infty\;,\hskip.3cm0<q<\infty\;.  \en
Here, the fiducial vector $|\eta\>$ is chosen to satisfy several conditions, 
which, using the shorthand $\<(\,\cdot\,)\>\equiv \<\eta|(\,\cdot\,)|\eta\>$, 
are specifically given by
  \bn\<Q^{-1}\>\equiv C<\infty\;,\hskip1cm\<\one\>=1\;,\hskip.3cm \<Q\>=1\;,
\hskip.3cm\<D\>=0\;,\hskip.3cm\<P\>=0\;. \en
The first condition is required, while the remaining conditions are chosen 
for convenience.
 The coherent states also admit a resolution of unity \cite{ask2} expressed 
in the form
\bn\one=\int|p,q\>\<p,q|\,d\tau(p,q)\;,\hskip1cm d\tau(p,q)=dp\,dq/2\pi 
C\;,\en integrated over the half plane $\ir\times\ir^+$.

In particular, diagonalizing the self-adjoint operator $Q=
\tint_0^\infty x|x\>\<x|\,dx$ in terms of standard Dirac-normalized 
eigenvectors, leads to a 
representation for the coherent-state overlap given by
\bn&&\<p,q|r,s\>\equiv\<\eta|e^{i\ln(q)D}e^{-ipQ}\,e^{irQ}e^{-i\ln(s)D}\,
|\eta\>\no\\
&&\hskip1.68cm=(qs)^{-1/2}\,\int_0^\infty\eta(x/q)^*e^{-ix(p-r)}\eta(x/s)\,
dx\;,
\en
where the fiducial function $\eta(x)=\<x|\eta\>$ denotes the Schr\"odinger 
representation of the fiducial vector $|\eta\>$.
It is important to observe, for some suitable function $F$, that
\bn \<p,q|r,s\>=F(q,p-r,s)\;,  \en
namely, that
$p$ and $r$ universally enter in the form $p-r$. 
It is also clear that $\<p,q|r,s\>$ defines a {\it continuous, 
positive-definite function}, which, apart from the continuity, means that
   \bn\sum_{n,m=1}^N\alpha^*_n\alpha_m\,\<p_n,q_n|p_m,q_m\>\ge0 \en
for arbitrary complex $\{\alpha_n\}_{n=1}^N$ and real $\{p_n,q_n\}_{n=1}^N$ 
sequences, with $N<\infty$. The function (5) may be taken as the
{\it reproducing kernel for a reproducing kernel Hilbert space} \cite{aro}. 
Note that the information in $\<p,q|r,s\>$ is enough to recover $\eta(x)$ 
apart from an overall constant phase factor. Thus different fiducial 
functions (not related by a constant phase factor) generate distinct 
reproducing kernels. Since each reproducing kernel Hilbert space has one 
and only one reproducing kernel \cite{aro}, it follows for different 
$\eta(x)$ that the Hilbert space functional realizations are completely 
disjoint except for the zero element. Basic elements of a dense set of 
vectors in each such Hilbert space are given by continuous functions of 
the form
 \bn \psi(p,q)\equiv\sum_{n=1}^N\alpha_n\<p,q|p_n,q_n\> \en
defined for arbitrary complex $\{\alpha_n\}_{n=1}^N$, and real 
$\{p_n,q_n\}_{n=1}^N$ sequences, with $N<\infty$. Let a second such 
function be given by
 \bn \phi(p,q)\equiv\sum_{j=1}^J\beta_j\<p,q|{\o p}_j,{\o q}_j\> \en
defined for arbitrary complex $\{\beta_j\}_{j=1}^J$ and real $\{{\o p}_j,
{\o q}_j\}_{j=1}^J$ sequences, with $J<\infty$. The inner product of two 
such vectors is then {\it defined} \cite{aro} to be
\bn \<\psi|\phi\>\equiv(\!\!(\psi(\cdot,\cdot)~_{\jmath\!\!\jmath}
\,\phi(\cdot,\cdot))\!\!)\equiv \sum_{n=1}^N\sum_{j=1}^J\a^*_n\beta_j
\<p_n,q_n|{\o p}_j,{\o q}_j\>\;,  \en
which when $|\phi\>=|\psi\>$ is, by definition, nonnegative. The resultant 
pre-Hilbert space is completed to a (reproducing kernel) Hilbert space 
$\cal C$ by including all Cauchy sequences in the norm $\||\psi\>\|
\equiv\sqrt{\<\psi|\psi\>}$ as $N$ tends to infinity. Lastly, we note 
that the space of functions appropriate to  one reproducing kernel is 
{\it identical} to the space of functions appropriate to a second 
reproducing kernel that is just a constant multiple of the first 
reproducing kernel. This fact does not contradict the uniqueness of the 
reproducing kernel for each Hilbert space because strictly 
different inner products are assigned in the two cases. Of course, the 
foregoing discussion applies quite generally and is not limited to any one
sort of reproducing kernel.

When the states $|p,q\>$ form a set of coherent states---as we assume 
in the present case---the inner product has an {\it alternative 
representation} given by a local integral of the form
  \bn \<\psi|\phi\>=\int\psi(p,q)^*\phi(p,q)\,d\tau(p,q)\;, \en
expressed in terms of $\psi(p,q)\equiv\<p,q|\psi\>$ and $\phi(p,q)
\equiv\<p,q|\phi\>$.
It follows that this formula holds for all elements of the completed 
Hilbert space $\cal C$, and, moreover, every element of the so-completed 
space is a {\it  bounded and continuous function}, the collection of which 
forms a rather special closed subspace of $L^2(\ir^2,d\tau)$. 

Thanks to the coherent-state resolution of unity, it follows that the 
coherent state overlap function satisfies the integral equation
  \bn \<p'',q''|p',q'\>=\int\<p'',q''|p,q\>\<p,q|p',q'\>\,d\tau(p,q)\;, \en
a basic relation, which, if it was established as a first step for the 
continuous function $\<p'',q''|p',q'\>\;[=\<p',q'|p'',q''\>^*]$, 
guarantees the existence of a local integral representation for the inner 
product of two arbitrary elements in the associated reproducing kernel 
Hilbert space. Several useful properties follow from this reproducing 
property. For example, repeated use of the resolution of unity
leads to the fact that
  \bn \<p'',q''|p',q'\>=\lim_{L\ra\infty}\int\!\cdots\!\int \prod_{l=0}^L
\<p_{l+1},q_{l+1}|p_l,q_l\>\,\prod_{l=1}^L\,d\tau(p_l,q_l)\;,\en
in which we have identified $p'',q''=p_{L+1},q_{L+1}$ and $p',q'=p_0,q_0$.
In turn, making an (unjustified!) interchange of the (continuum) limit with 
the integrations, and writing for the integrand the form it would assume 
for continuous and differentiable paths, gives rise to the suggestive but 
strictly formal expression \cite{klbo}
  \bn \<p'',q''|p',q'\>=\int e^{-i\tint_0^T q(t)\,{\dot p}(t)\,dt}\,
\D\tau(p,q)\;, \en
which determines a formal path integral representation for the kinematics 
that applies for any $T>0$. Thus, the existence of a coherent state 
resolution of unity is the necessary condition to introduce a traditional 
coherent state phase-space path integral representation for the kinematics, 
which specifically leads to the reproducing kernel. A path integral for the 
kinematics is a necessary prerequisite to obtain a path integral for the 
dynamics. While less than ideal, the formal path integral itself may be 
used as a starting point for quantization. Even though the formal nature of 
the path integral renders it basically undefined, one may always 
(re)introduce a regularization by a lattice-limit formulation (as above), 
using suitable ingenuity to choose an acceptable integrand. This procedure 
is more or less standard by now.

However, it must be appreciated that reproducing kernels for reproducing 
kernel Hilbert spaces are {\it not required} to fulfill a (positive) 
local integral 
representation for the inner product. In cases where the integral for the 
resolution of unity does not exist, one must accept the inner product that 
is given directly by the reproducing kernel, which means that the integral 
relation (12) is replaced by
\bn(\!\!(\<\cdot,\cdot|p'',q''\>~_{\jmath\!\!\jmath}\,\<\cdot,
\cdot|p',q'\>)\!\!)\equiv\<p'',q''|p',q'\>\;.  \en
When this is the case we say that $\{|p,q\>\}$ forms a set of 
{\it weak coherent states} \cite{klbo1}, i.e., the elements of 
$\{|p,q\>\}$ span the 
Hilbert space $\frak H$, but do not admit a local integral representation 
for the inner product of elements in the associated reproducing kernel 
Hilbert space.

A simple  example of a reproducing kernel Hilbert space without a local 
integral representation for the inner product is determined, for $u'',u'
\in\ir$, by the reproducing kernel $\<u''|u'\>\equiv\exp[-(u''-u')^2]$; 
here, one must use $(\!\!(\<\,\cdot\,|u''\>~_{\jmath\!\!\jmath}\,\<\,
\cdot\,|u'\>)\!\!)\equiv\<u''|u'\>$, which is then extended by linearity 
and continuity to all Hilbert space vectors. 

A more relevant set of examples is given by the following discussion 
applied to our simple model. Let $\a>-1/2$, and choose $\eta(x)\equiv N\,
x^\a\,\exp(-\beta x)$. Here the factor $N$ is fixed by requiring 
$\tint_0^\infty|\eta(x)|^2\,dx=1$. The two conditions $\<Q\>=1$ and 
$\<Q^{-1}\>=C<\infty$, lead to $\beta-\half=\a>0$; in this case 
$C=1-1/(2\beta)$. 
In turn, the reproducing kernel is given explicitly \cite{affi} by
\bn &&\<p,q|r,s\>=\bigg[\frac{(qs)^{-1/2}}{\half(q^{-1}+s^{-1})+
i\half\beta^{-1}(p-r)}\bigg]^{2\beta}\no\\
&&\hskip1.73cm=\exp(\!\!(-2\beta\ln\{[\half(q^{-1}+s^{-1})+
i\half\beta^{-1}(p-r)]/(qs)^{-1/2}\})\!\!)\;;\no\\&&\en
the second form is given for  comparison purposes to the gravitational case.
As long as $\beta>\half$ it follows that the states $|p,q\>$ are a set of 
coherent states with a proper resolution of unity and therefore a local 
integral representation for the inner product exists. In this case, path 
integrals exist as lattice limits, and the whole situation seems familiar. 
On the other hand, if $0<\beta\le\half$, the overlap function 
$\<p,q|r,s\>$ defined above is still a positive-definite function and, 
therefore, it is a valid reproducing kernel which leads to an associated 
reproducing kernel Hilbert space; however, {\it such a Hilbert space 
does not admit a local integral representation for the inner product in 
terms of the given representatives}. Therefore, there is {\it no} 
conventional coherent state path integral for the kinematics, and thus 
also for the dynamics, in a reproducing kernel Hilbert space representation 
when $0<\beta\le\half$. This lack of a conventional coherent state path 
integral representation may appear to be detrimental to any program to 
introduce quantization, dynamics, 
etc.---but there is hope. 

There is another way to generate the reproducing kernel for the given 
family of fiducial vectors which may be applied for all $\beta>0$. Let us 
first focus on $\beta>\half$. In that case, observe, by 
construction and using $\d_p\equiv\d/\d p$, etc.,
 that for every $|\psi\>\in{\frak H}$,
 \bn B\,\psi(p,q)\equiv \{-iq^{-1}\d_p+1+\beta^{-1}q\,\d_q\}\,\psi(p,q)=0\;, 
\en
an equation which represents a (complex) {\it polarization} \cite{wood} of 
$L^2(\ir^2,d\tau)$.
It follows that the second-order differential operator $A\equiv\half\beta 
B^\dag B\ge0$, and therefore $A$ can be used to generate a semigroup. In 
particular, for any $T>0$ and as $\nu\ra\infty$, the expression
  $e^{-\nu TA}$ becomes a {\it projection operator} onto the subspace 
$\cal C$ of solutions to the polarization equation \cite{dkp}. In a two 
degree of freedom Schr\"odinger representation---symbolized by $|p,q)$, 
where $(p,q)\in\ir\times\ir^+$ and $(p,q|r,s)=\delta(p-r)\delta(q-s)$---it 
follows, from a two-variable Feynman-Kac-Stratonovich path integral 
formula \cite{roe},  that
  \bn &&\hskip-.3cm\<p'',q''|p',q'\>\equiv\lim_{\nu\ra\infty}(p'',q''|
e^{-\nu TA}|p',q')\no\\
 &&\hskip-1cm=\lim_{\nu\ra\infty}{\cal N}\int e^{-i\tint_0^T q(t)\,
{\dot p}(t)\,dt-(1/2\nu)\tint_0^T[\beta^{-1} q(t)^{2}{\dot p}(t)^2+\beta 
q(t)^{-2}{\dot q}(t)^2]\,dt}\,\D p\,\D q\no\\
 &&\hskip-1cm\equiv\lim_{\nu\ra\infty}e^{\nu T/2}\int e^{-i\tint_0^T q(t)
\,dp(t)}\,dW^\nu(p,q)\;, \en
where $W^\nu$ denotes a two-dimensional Wiener measure with diffusion 
constant $\nu$, pinned at $t=0$ to $p',q'$ and at time $t=T$ to $p'',q''$, 
which is supported on a space of constant negative curvature $R=-2/\beta$. 
It is noteworthy in (18) that the variable $p(t)$ enters only in the form 
${\dot p}(t)$, a fact which leads to the result depending only on the 
difference, $p''-p'$. For every $\nu<\infty$, and with probability one, 
all Wiener paths in the given path integral are {\it continuous}, and, 
for purposes of coordinate transformations, it is convenient to adopt the 
(midpoint) Stratonovich rule to define the stochastic integral 
$-\tint q(t)\,dp(t)$ since, in that case, the rules of the ordinary 
calculus hold. Such a representation is said to involve a 
{\it continuous-time regularization} \cite{kla6}.

Now we consider the case where $0<\beta\le\half$. The solutions to (17) 
are, up to a factor,  analytic functions, but they are no longer square 
integrable, as is clear from the fact that the large $q$ behavior of (16) is 
controlled only by the factor $q^{-\beta}$. As a consequence, the operator 
$A\ge0$ has only a continuous spectrum. The family of operators 
$e^{-\nu T A}$ is still a semigroup, and the expression
$(p'',q'|e^{-\nu T A}|p',q')$ still has the formal path integral 
representation 
given in
the middle line of (18). However, as $\nu\ra\infty$, $e^{-\nu T A}$ does not
lead to a projection operator; instead we need to extract the germ of that 
semigroup as $\nu\ra\infty$. If we let $E\ge0$ denote continuum eigenvalues 
for the operator $\half B^\dag B$, we can write 
  \bn (p'',q''|e^{-\nu TA}|p',q')=\int (p'',q''|E,v)\,e^{-\nu T\beta E}
(E,v|p',q')
\,\rho(E,v)\,dE\,dv\;, \en
for some density of states $\rho(E,v)$. Here, the variable $v$ labels 
degeneracy for $\half B^\dag B$. For the sake of illustration, let 
us assume that $\rho(E,v)\simeq {\o C}E^w\,{\o\rho}(v)$, $w>-1$, for $E\ll1$. 
We base this 
assumption on the fact that there is no 
reason for an $E$-dependent degeneracy for very tiny $E$. Thus, we
consider the expression 
 $J(\nu)\equiv(\nu\beta T)^{w+1}/{\o C}\,\Gamma(w+1)$, and are led to the 
fact  that
 \bn &&\hskip-.5cm\lim_{\nu\ra\infty}\,J(\nu)\,(p'',q''|e^{-\nu TA}|p',q')
\no\\
&&\lim_{\nu\ra\infty}\int (p'',q''|E,v)\,e^{-\nu T\beta E}(E,v|p',q')
\,J(\nu)\,\rho(E,v)\,dE\no\\
&&\hskip.5cm=\int (p'',q''|0,v)(0,v|p',q')\,{\o\rho}(v)\,dv\;. \en
In effect, this procedure has enabled us to pass to the germ of the 
semigroup.  Observe 
that the rescaling factor is independent 
of the coherent state labels, and thus we are only making a $\nu$-dependent 
rescaling before the limit $\nu\ra\infty$ is taken. If necessary, we can 
rescale our expression by letting $J(\nu)\ra{\o J}(\nu)=
M(p'',q'')\,M(p',q')\,J(\nu)$ to achieve normalization without effecting 
its positive-definite character. In summary, for 
$0<\beta\le\half$, we claim that instead of (18) we can write 
\bn &&\hskip-.3cm\<p'',q''|p',q'\>\equiv\lim_{\nu\ra\infty}\,{\o J}(\nu)
\,(p'',q''|
e^{-\nu TA}|p',q')\no\\
 &&\hskip-1cm=\lim_{\nu\ra\infty}{\cal{\o N}}\int e^{-i\tint_0^T q(t)\,
{\dot p}(t)\,dt-(1/2\nu)\tint_0^T[\beta^{-1} q(t)^{2}{\dot p}(t)^2+\beta 
q(t)^{-2}{\dot q}(t)^2]\,dt}\,\D p\,\D q\no\\
 &&\hskip-1cm\equiv\lim_{\nu\ra\infty}{\o J}(\nu)\,e^{\nu T/2}
\int e^{-i\tint_0^T 
q(t)
\,dp(t)}\,dW^\nu(p,q)\;, \en
Convergence in this case is initially regarded in the sense of distributions. 
To lead to the desired result, we appeal to analyticity (up to a specific 
factor) of 
the result in $q^{-1}+i\beta^{-1}p$, analyticity (up to another
factor) in the variable $s^{-1}-i\beta^{-1}r$, and dependence on $p-r$. 
 Note that ${\o J}(\nu)$ can always be determined self 
consistently by insisting that $\<p,q|p,q\>=1$ for all $(p,q)$.
In simpler terms, we can always regard ${\o J}(\nu)$ as part of the needed 
normalization coded into $\cal{\o N}$ in the formal path integral expression.

As will become evident later, various features of reproducing kernels, 
reproducing kernel Hilbert spaces, and associated rules for defining inner 
products illustrated above will carry over into the quantum gravity case as 
well.
\subsection{Operators and symbols}
In addition to the properties of the reproducing kernel Hilbert space, 
certain {\it symbols} associated with operators are important. Let us 
introduce the upper symbol $H(p,q)$ associated to the operator $\H(P,Q)$ 
and defined, modulo suitable domain conditions, by the expression
 \bn H(p,q)\equiv\<p,q|\H(P,Q)|p,q\>=\<\H(p+P/q,qQ)\>\;.\en
For example, if $\H(P,Q)=P^2-Q^{-1}$ denotes a quantum Hamiltonian, then 
$H(p,q)= p^2+\<P^2\>/q^2-C/q$.
Since $C=O(1)$ (e.g., for large $\beta$) and $\<Q\>=1$, it follows that 
$\<P^2\>=O(\hbar^2)$. Observe that $H$ basically agrees with the expected 
classical Hamiltonian in the limit that $\hbar\ra0$, but prior to that 
limit $H$ includes a quantum induced barrier to singularities in solutions 
of the usual classical equations of motion. We adopt the expression 
$H(p,q)$ as the ($\hbar$-augmented) classical Hamiltonian and refer to the 
connection between the quantum generator $\H$ and the classical generator 
$H$ as the {\it weak correspondence principle} \cite{kl9}. In this way, the 
classical and quantum theories may both {\it coexist}, as they do in Nature. 

There is also another set of symbols that are important. We introduce the 
lower symbol $h(p,q)$ which is related to the operator $\H(P,Q)$ by the 
relation
  \bn \H(P,Q)=\int h(p,q)\,|p,q\>\<p,q|\,d\tau(p,q)\;. \en
For the one-parameter class of fiducial vectors leading to (16), it follows 
that a dense set of operators admit such a symbol for a reasonable set of 
functions. 

In the quantum gravity case, there are twin goals: (i) to ensure that the 
field operators of interest are well defined and locally self adjoint in 
the given field operator representation; and (ii) to choose locally self 
adjoint constraint operators that have a weak correspondence principle 
which connects them with the desired form of the classical constraint 
generators (possibly $\hbar$ augmented).

\subsection{Imposition of constraints}
We adopt the projection operator approach to the quantization of systems 
with constraints \cite{kla1,sha,gov,kla2}. Let $\{\Phi_\a(P,Q)\}_{\a=1}^A$, 
$A<\infty$, denote a set of constraints each given by a self-adjoint 
operator. Further assume that
$\Phi\cdot\Phi\equiv\Sigma_{\a=1}^A(\Phi_\a)^2$ is also self adjoint. 
We define the (provisional) physical Hilbert space ${\frak H}_{\rm\,phys}
\equiv\E{\frak H}$, where $\E=\E^\dagger=\E^2$ is a uniquely defined 
projection operator, and in turn choose
  \bn \E\equiv\E(\!\!(\Phi\cdot\Phi\le\delta(\hbar)^2)\!\!)\;,  \en
where $\delta(\hbar)$ is {\it not} a $\delta$-function but a small, 
positive, possibly $\hbar$ dependent, {\it regularization parameter} 
for the set of constraints. As shown below, $\delta(\hbar)$ is chosen so 
that $\E$ is the desired projection operator. This choice may entail a 
specific representation of the Hilbert space and a suitable limit as 
$\delta\ra0$ to extract the germ of the projection operator. 

We may illustrate this latter situation for the constraint $\Phi\equiv Q-1$, 
assuming initially that $\delta<1$. In this case 
\bn \<\psi|\E(\!\!((Q-1)^2\le\delta^2)\!\!)|\phi\>=\int_{1-\delta}^{1+\delta} 
dx\int d\sigma(y)\,\psi(x,y)^*\phi(x,y)\;, \en
where $\sigma$ accounts for any degeneracy that may be present. 
When restricted to functions $\psi_o$ and $\phi_o$ in the dense set 
$\frak D$, where 
    \bn{\frak D}\equiv\{{\rm\,polynomial}\,(x,y)\,e^{-x^2-y^2}\}\;,\en
and rescaled by a suitable factor,
the projection operator matrix elements lead to the expression
\bn&&(2\delta)^{-1}\<\psi_o|\E(\!\!((Q-1)^2\le\delta^2)\!\!)|\phi_o\>\no\\
&&\hskip2cm=(2\delta)^{-1}\int_{1-\delta}^{1+\delta} dx\int d\sigma(y)\,
\psi_o(x,y)^*\phi_o(x,y)\;. \en
Now, as $\delta\ra0$, this expression becomes
 \bn \int\psi_o(1,y)^*\phi_o(1,y)\,d\sigma(y)\equiv(\!(\psi_o,\phi_o)\!)\;. 
\en
Interpreting this final expression as a sequilinear form, one completes the 
desired Hilbert space by adding all Cauchy sequences in the associated norm 
$\||\psi_o)\!)\|\equiv\sqrt{(\!(\psi_o,\psi_o)\!)}\,$. The result is the true 
physical Hilbert space in which the constraint $Q-1=0$ is fulfilled, and 
this example illustrates how constraints are to be treated when $\Phi\cdot
\Phi$ has its zero in the continuous spectrum.

A second example of an imposition of constraints is given by $\Phi_1=Q-1$, 
as before, along with $\Phi_2=D$. This situation corresponds to second 
class constraints, and serves as a simple qualitative model of what occurs 
in the gravitational case. In this case
  \bn \E=\E(\!\!(D^2+(Q-1)^2\le \delta(\hbar)^2)\!\!) \;. \en
Here, the left-hand side of the argument can be regarded as a 
``Hamiltonian'', and the ground state $|0'\>$ for such a system 
(nondegenerate, in the present example) can be sought. In particular, 
there are two positive parameters, $\delta'$ and $\delta''$, such that, 
for all $\delta$ with $\delta'\le\delta<\delta''$, then 
\bn \E=\E(\!\!(D^2+(Q-1)^2\le \delta(\hbar)^2)\!\!)\equiv |0'\>\<0'| \;. \en
This is the desired choice to make for $\E$ in the case where $\Phi\cdot
\Phi$ has a discrete spectrum near zero that does not include zero.

For completeness, if $\Phi\cdot\Phi$ has a discrete spectrum including 
zero, then it suffices to choose
  \bn \E=\E(\!\!(\Phi\cdot\Phi=0 )\!\!)=\E(\!\!(\Phi\cdot\Phi\le\delta
(\hbar)^2 )\!\!)\;, \en
where in the present case $\delta(\hbar)>0$ is chosen small enough to 
include only the subspace for which $\Phi\cdot\Phi=0$. This simple case 
does not seem to arise in quantum gravity. See \cite{kla1, kla2} for a 
discussion of gauge invariance.
  
When the number of constraints is infinite, $A=\infty$, as will be the 
case for a field theory, then a slightly different approach is appropriate. 
One form this takes is dealt with in Sec.~4.

\subsection{Appearance of time}
In a reparametrization invariant problem in quantum mechanics it is typical 
that dynamics is cast in the guise of kinematics at the expense of 
introducing an additional degree of freedom plus a first-class constraint; 
see, e.g., \cite{gov2}. Let the resultant kinematical reproducing kernel 
with the extra degree of freedom be given by $\<p'',q'',s'',t''|
\E|p',q',s',t'\>$, where $\E$ is the projection operator enforcing 
the constraint. Next, reduce this expression, for example, as in the 
procedure
  \bn \<p'',q'',t''|p',q',t'\>\equiv \tint\tint\<p'',q'',s'',t''|
\E|p',q',s',t'\>\,ds''\,ds'\;. \en
The result is a new positive-definite function that can be used to define 
a reproducing kernel Hilbert space. However, it may well happen that the 
following identity holds
  \bn &&\<p'',q'',t''|p',q',t'\>=
(\!\!(\<\cdot,\cdot,\cdot|p'',q'',t''\>~_{\jmath\!\!\jmath}\, 
\<\cdot,\cdot,\cdot|p',q',t'\>)\!\!)\no\\
  &&\hskip3.25cm=(\!\!(\<\cdot,\cdot,t|p'',q'',t''\>~_{ \jmath\!\!\jmath}
\, \<\cdot,
\cdot,t|p',q',t'\>)\!\!)\;.  \en
This equation means that the space spanned by the states $|p,q,t\>$ by 
varying $p,q$ {\it and} $t$ is the {\it same space} spanned, by the same 
states, by varying $p$ and $q$ but with $t$ held {\it fixed} at some 
value (e.g., $t=0$). This situation implies that the states $|p,q,t\>$ are 
{\it extended coherent states} in the sense of \cite{whi}, and, 
in particular, thanks to using canonical group coordinates in the 
coherent-state parameterization, that $|p,q,t\>=\exp{(-i\H t)}|p,q,0\>$ 
for some self-adjoint ``Hamiltonian'' $\H$. The parameter $t$ is then 
recognized as the ``time'' $t$. For an explicit example of how this 
procedure works in detail see \cite{klaa}. 

It is expected that a suitable time parameter will emerge in the extension 
of these ideas to the gravitational case.
\subsection{Matrix generalization}
Our preceding analysis has been confined to a single $p$ and $q$ and the 
associated affine quantum operators. In any generalization to the 
gravitational case, it will first be necessary to generalize the 
preceding discussion to $3\times3$ (or more generally to $s\times s$) 
matrix degrees of freedom and repeat an analysis similar to that of the 
present section. We do not include this discussion here since that is the 
subject of a separate work \cite{wat}. It is safe to 
say, apart from some technical details, that there are no special 
surprises in this generalization, and the basic concepts that we shall 
need are already present in the simplest case on which we have concentrated.  

\section{Gravitational Kinematics}
\subsection{Preliminaries}
Let us start with the introduction of a three-dimensional {\it topological 
space} $\cal S$ which locally is isomorphic to a subset of $\ir^3$. Locally, 
we generally use three ``spatial" coordinates, say $x^j$, $j=1,2,3$, to 
label a point in $\cal S$. This labeling is nonsingular and thus one-to-one. 
Whether a single coordinate chart covers $\cal S$ depends on the global 
topological structure of $\cal S$. Let us fix this global topological 
structure from the outset---for example, topologically equivalent to $\ir^3$, 
$S^3$, $T^3$, etc. The theory of quantum gravity developed here does not 
engender topological changes of the underlying topological space $\cal S$.
Note this lack of topological change applies only to the space $\cal S$. It
is unrelated to any presumed ``space'' and/or  ``topology'' associated 
with any 
quantum metric tensor, which, after all, is typically distributional 
in character.

If the space $\cal S$ is such that more than one coordinate patch is 
required  we arrange for the necessary matching conditions and rename the 
coordinates within each patch by $x^j$, $j=1,2,3$, for some domain. We can 
also consider alternative coordinates, say ${\o x}^j$, $j=1,2,3$, which are 
also nonsingular. We admit only differentiable coordinate transformations 
such that the Jacobian $[\d x/\d {\o x}]\equiv \det(\d x^j/\d {\o x}^k)\ne 0$ 
everywhere. The group composed of such invertible coordinate transformations 
is the {\it diffeomorphism group}.

We can also introduce functions on the space $\cal S$ which in coordinate 
form may be denoted by $f(x)$. A scalar function is one for which 
${\o f}({\o x})=f(x)$, while a scalar density of weight one satisfies
${\o b}({\o x})=[\d x/\d {\o x}]\,b(x)$, or stated as a volume form, 
$dV={\o b}({\o x})\,d^3\!{\o x}=b(x)\,d^3\!x$. Observe that it is not 
necessary to have a metric in order to have a volume form. Whether 
$\cal S$ is compact or noncompact, we assume that $0<b(x)<\infty$ for 
all $x$ and therefore $0<b(x)^{-1}<\infty$ for all $x$ as well. These 
properties are still valid after a nonsingular coordinate change. Integrals 
of a scalar function take the form $\int f\,dV=\int f(x)\,b(x)\,d^3\!x$ 
and are {\it invariant} under any coordinate transformation in the 
diffeomorphism group.

In the ADM (Arnowitt, Deser, Misner) \cite{adm} canonical formulation of 
classical gravity there are two fundamental fields $g_{kl}(x)\;[=g_{lk}(x)]$ 
and $\pi^{kl}(x)\;[=\pi^{lk}(x)]$. The metric $g_{kl}(x)$ transforms as a 
(two-valent covariant) tensor, while the momentum $\pi^{kl}(x)$ transforms 
as a (two-valent contravariant) tensor density of weight one. Thus  
$\int g_{kl}(x)\,\pi^{kl}(x)\,d^3\!x$ [or even $\int g_{kl}(x,t)\,
{\dot\pi}^{kl}(x,t)\,d^3\!x\,dt$] is an invariant under diffeomorphism 
group transformations (on the spatial hyperspace, of course). Note well: 
The latter example pertains to a 
generalization of the former one including an additional independent 
variable $t$ which possibly could be identified with (coordinate) ``time".

A metric is not an arbitrary tensor but is restricted to be positive 
definite. Specifically, for any real $\alpha^j$, $j=1,2,3$, where 
$\Sigma_{j=1}^3(\alpha^j)^2>0$, it follows that $\alpha^k\,g_{kl}(x)\,
\alpha^l>0$ for all $x$. As a consequence, the positive-definite 
(two-valent contravariant) tensor $g^{kl}(x)$ exists at each point and 
is defined so that $g^{kl}(x)\,g_{lm}(x)=\delta^k_m$. In addition, 
$\sqrt{g(x)}\equiv\sqrt{\det[g_{kl}(x)]}>0$ transforms as a scalar 
density of weight one. Thus, as is well known, $\sqrt{g(x)}\,d^3\!x$ 
characterizes a volume form, but this choice ties the volume form to a 
specific metric, or at least to a specific class of metrics. This close 
association to specific metrics is something we would like to avoid, and 
it leads us to choose $b(x)\,d^3\!x$ as the preferred volume form. 
Of course, if $b(x)=\sqrt{g(x)}$ everywhere in any coordinate system, 
then the volume form $b(x)\,d^3\!x$ is identical to the one based on a 
metric space and given by $\sqrt{g(x)}\,d^3\!x$.

\subsection{Reproducing kernel---original Hilbert space}
A study of canonical quantum gravity begins with the introduction of metric 
and momentum local quantum field operators,  which we denote by 
$\sigma_{kl}(x)\;[=\sigma_{lk}(x)]$ and $\mu^{kl}(x)\;[=\mu^{lk}(x)]$, 
respectively. For such fields one postulates  the canonical commutation 
relations
  \bn &&[\sigma_{kl}(x),\,\mu^{rs}(y)]=i\,\delta^{rs}_{kl}\,\delta(x,y)\;, 
\no\\
    &&[\sigma_{kl}(x),\,\sigma_{rs}(y)]=0\;,\no\\
    &&[\mu^{kl}(x),\,\mu^{rs}(y)]=0 \;,\en
with $\delta^{rs}_{kl}\equiv (\delta^r_k\delta^s_l+\delta^r_l\delta^s_k)/2$. 
Since the right-hand side of the first equation is a tensor density of 
weight one, it is consistent that we define $\sigma_{kl}(x)$ to be a 
tensor and $\mu^{rs}(x)$ to be a tensor density of weight one. However, 
just as its one-dimensional counterpart, there are no local self-adjoint 
field and momentum operators that satisfy the canonical commutation 
relations as well as the requirement that $\{\sigma_{kl}(x)\}>0$. To 
arrive at a suitable substitute set of commutation relations, we introduce, 
along with the local metric field operator $\sigma_{kl}(x)$, the 
local ``scale'' field operator $\kappa^r_k(x)$ which together obey the 
{\it affine commutation relations} \cite{kla,ish,pli}
 \bn  &&[\kappa^r_k(x),\,\kappa^s_l(y)]=i\,\half\,[\delta^s_k\,\kappa^r_l(x)-
\delta^r_l\,\kappa^s_k(x)]\,\delta(x,y)\;,\no\\
    &&[\sigma_{kl}(x),\,\kappa^r_s(y)]=i\,\half\,[\delta^r_k\,
\sigma_{ls}(x)+\delta^r_l\,\sigma_{ks}(x)]\,\delta(x,y)\;,\no\\
  &&[\sigma_{kl}(x),\,\sigma_{rs}(y)]=0\;. \en
In these relations, $\sigma_{kl}(x)$ remains a tensor, while $\kappa^r_s(x)$ 
is a tensor density of weight one under coordinate transformations.
The local operators $\kappa^r_s(x)$ are generators of the GL(3,R)$^\infty$ 
group \cite{ish,pli}, while the local operators $\sigma_{kl}(x)$ are 
commuting ``translations'' coupled with the GL(3,R)$^\infty$ group by a 
semi-direct product. The given affine commutation relations are the natural 
generalization of the one-dimensional affine commutation relation presented 
in (1). In the case of one degree of freedom, the affine commutation
relations follow from the canonical ones, while in the case of fields this 
is, strictly speaking, incorrect. It is true that
  \bn\kappa^r_k(x)=\half[\sigma_{kl}(x)\,\mu^{lr}(x)+\mu^{rl}(x)
\sigma_{lk}(x)]_R \;,\en
where the subscript $R$ denotes an infinite multiplicative renormalized 
product to be defined later. However, the presence of an infinite 
rescaling means that either the canonical or the affine set of commutation 
relations can hold, but not both at the same time. Since it is the affine 
commutation relations that are consistent with local self-adjoint operator 
solutions enjoying metric positivity, we shall adopt the noncanonical 
affine commutation relations. The choice of the affine commutation relations 
means that the canonical commutation relations do {\it not} hold, and 
therefore we are dealing with a {\it non}canonical quantization of the 
gravitational field. 

Accepting the affine field operators as generators, we introduce a primary 
set of normalized affine coherent states each of which---in a deliberate
abuse of notation---is defined by 
  \bn|\pi,g\>\equiv e^{i\tint \pi^{kl}(x)\,\sigma_{kl}(x)\,d^3\!x}\,e^{-i
\tint\gamma^r_s(x)\,\kappa^s_r(x)\,d^3\!x}\,|\eta\>\;,  \en
for a suitable fiducial vector $|\eta\>$ characterized below.
In $|\pi,g\>$ the argument ``$\pi$'' denotes the momentum matrix field 
$\pi^{ab}$ while ``$g$'' denotes the metric matrix field $g_{ab}$.
By all rights, the states in question should have been called 
$|\pi,\gamma\>$, but as
we shall see, the overlap of two such states, {\it for the featured choice of}
$|\eta\>$ [cf. (70)],
depends only on the matrix (for each point $x$) $g\equiv\exp(\gamma^T/2)
\exp(\gamma/2)\equiv\{g_{ab}\}$, where $T$ means 
``transpose''.\footnote{Observe by this parametrization that $g>0$ as 
opposed to a 
traditional triad for which $g\ge0$. To see that this distinction may 
possibly make a real difference see \cite{affi}. We remark that the 
nonsymmetric matrix $\exp(\gamma/2)$ would have  relevance for spinor 
fields.}
If the space $\cal S$
is noncompact, then, as smooth $c$-number fields, both $\pi$ and $\gamma$ 
should go to zero sufficiently 
fast so that the indicated smeared field operators are indeed self-adjoint 
operators and generate unitary transformations as required. On the other 
hand, as we shall shortly see, this asymptotic behavior can, effectively, 
be significantly relaxed.

The overlap of two such coherent states leads to an expression of the form
  \bn\<\pi'',g''|\pi',g'\>=F(g'',\pi''-\pi',g') \en
for some continuous functional $F$ which depends only on the difference of 
the fields, $\pi''(x)-\pi'(x)$, an analog of which already occurred for the 
one-dimensional example. Whatever choice is made for the fiducial vector 
$|\eta\>$, the coherent state overlap function defines a continuous, 
positive-definite functional which, therefore, defines a reproducing kernel 
and its associated (separable) reproducing kernel Hilbert space $\cal C$. 
By construction, therefore, the set of coherent states $\{|\pi,g\>\}$ span 
the Hilbert space $\frak H$. As such they form a basis (overcomplete to 
be sure!) for $\frak H$. Based on arguments to follow, we are led to the 
proposal [cf., (16)] that
  \bn &&\<\pi'',g''|\pi',g'\>=\no\\
&&\hskip-.7cm\exp\bigg(\!-\!2\int b(x)\,d^3\!x\,\ln\bigg\{  \frac{
\det\{\half[g''^{kl}(x) +g'^{kl}(x)]+i\half b(x)^{-1}[\pi''^{kl}(x)-
\pi'^{kl}(x)]\}} {(\det[g''^{kl}(x)])^{1/2}\,(\det[g'^{kl}(x)])^{1/2}}  
\bigg\}\bigg) \no\\
&&\en
{\it This equation is central to our analysis of quantum gravity.}

The coherent-state overlap (39) may be read in two qualitatively different 
ways. Although arrived at on the basis that $\pi''(x),\pi'(x)\ra0$ and 
$\gamma''(x),\gamma'(x)\ra0$ as $|x|\ra\infty$, the given expression exists 
for a far wider limiting behavior. In particular, suppose there is a 
{\it fixed asymptotic behavior} such that for both $\pi=\pi''$ and 
$\pi=\pi'$, $\pi^{ab}(x)-{\tilde\pi}^{ab}(x)\ra0$ and for both $g=g''$ and 
$g=g'$, $g_{ab}(x)-{\tilde g}_{ab}(x)\ra0$, all terms vanishing sufficiently 
fast as $|x|\ra\infty$. In this case $g\equiv\exp(\gamma^T/2){\tilde g}
\exp(\gamma/2)$. Note that the asymptotic fields can depend on $x$. 
In this case the coherent-state overlap still
holds in the form given. This kind of asymptotic behavior reflects a change 
of the fiducial vector $|\eta\>$, which now depends on the explicitly 
chosen asymptotic form for the momentum and metric---or, equivalently, as 
we effectively do, one can hold $|\eta\>$ fixed and change the 
representations of the operator [cf., (71)]. By choosing a suitable asymptotic
momentum and metric one can, in effect, redefine the topology of the 
underlying space $\cal S$. However, for simplicity, we shall assume simple 
Euclidean-like asymptotic behavior of the momentum and metric [
${\tilde\pi}^{ab}(x)\equiv 0$ and  ${\tilde\gamma}^r_s(x)\equiv0$, i.e., 
${\tilde g}_{ab}(x)\equiv\delta_{ab}$].  

A second way to study the coherent-state overlap is under coordinate 
transformations. 
Observe that $\<\pi'',g''|\pi',g'\>$ is {\it invariant} if, everywhere, 
we make the replacements \vskip.2cm 
\hskip.5cm(i) $b(x)\,d^3\!x$ \hskip.3cm by\hskip.3cm ${\o b}({\o x})\,d^3
\!{\o x}=b(x)\,d^3\!x$, \vskip.1cm
\hskip.5cm(ii) $g^{kl}(x)$ \hskip.3cm by\hskip.3cm ${\o g}^{kl}({\o x})=
M_{\;r}^{k}(x)\,g^{rs}(x)\,M_{\;s}^{l}(x)$,  \vskip.1cm
\hskip.5cm(iii) $b(x)^{-1}\pi^{kl}(x)$ \hskip.3cm by\hskip.3cm 
${\o b}({\o x})^{-1}{\o\pi}^{kl}({\o x})=b(x)^{-1}M_{\;r}^{k}(x)\,
\pi^{rs}(x)\,M_{\;s}^{l}(x)$, \vskip.2cm
\noindent all for an arbitrary nonsingular matrix $M\equiv\{M_{\;r}^{k}\}$, 
$M_{\;r}^{k}(x)\equiv (\d{\o x}^k /\d x^r)(x)$, that arises from a 
nonsingular coordinate transformation $x\ra{\o x}={\o x}(x)$. It suffices to 
restrict attention to those coordinate transformations continuously connected 
to the identity. When $\cal S$ is compact, a wide class of  $M$ is allowed; 
when $\cal S$ is noncompact, the allowed elements $M$ must also map coherent 
states into coherent states for the same fiducial vector. This restriction  
excludes any connection by coordinate transformations of two field sets with 
fundamentally different asymptotic behavior; such fields live in disjoint 
sets. The invariance under admissible coordinate transformations is 
symbolized by the statement that
  \bn \<{\o\pi}'',{\o g}''|{\o\pi}',{\o g}'\>=\<\pi'',g''|\pi',g'\>\en
for all suitable $M$. Since the allowed $M$ form a representation of the 
connected component of the diffeomorphism group, it follows from this 
identity, and suitable continuity, that for sufficiently restricted $M$, the 
transformation $|\pi,g\>\ra|{\o\pi},{\o g}\>$ is induced by a {\it unitary 
transformation}, specifically that
  \bn &&|{\o\pi},{\o g}\>\equiv U(M)\,|\pi,g\>\;,\\
    &&\hskip-.15cm U(M)\equiv \exp[-i\tint N^j(x)\,\H_j(x)\,d^3\!x]\;. \en
Here $\H_j(x)$ denotes a local operator tensor density of weight one while 
$N^j(x)$ denotes a $c$-number tensor with sufficiently rapid decay at 
spatial infinity.  Furthermore, using the shorthand  that  $\tint N^jH_j
\equiv\tint N^j(y)H_j(y)\,d^3\!y$, the connection between $M$ and $N^j$ is 
implicitly given by
\bn && M^{a}_{\; r}g^{rs}M^{b}_{\;s}=g^{ab}-\{\tint N^jH_j,\,g^{ab}\}
  +\frac{1}{2!}\{\tint N^kH_k,\,\{\tint N^jH_j,\,g^{ab}\}\}+...\no\\
   &&\hskip1.9cm \equiv e^{-\{\tint N^jH_j,\;\cdot\;\}}\,g^{ab}\;, \en
where $\{\,\cdot\,,\,\cdot\,\}$ denotes the classical Poisson brackets, and 
specifically, e.g., $\{g_{ab}(x),\,\pi^{rs}(y)\}=\delta_{ab}^{rs}\,
\delta(x,y) $.
In this expression, $H_j(x)=-g_{jk}(x)\pi^{kl}_{\;\;\;|l}(x)$,  $j=1,2,3$, 
where $({~})_{|l}$ is the covariant derivative with respect to the 
$3\times3$ metric, denotes the classical generators of the diffeomorphism 
group \cite{mtw}. The relationship of $H_j(y)$ and $\H_j(y)$ may be 
determined as follows. 
Expansion of the relation
 \bn \<\pi'',g''|\,e^{-i\tint N^j(x)\H_j(x)\,d^3\!x}\,|\pi',g'\>=\<\pi'',
g''|{\o\pi}',{\o g}'\>  \en
to first order in $N^j$ leads to 
   \bn &&\hskip-.5cm\<\pi'',g''|\tint N^j(x)\H_j(x)\,d^3\!x\,|\pi',g'\>/
\<\pi'',g''|\pi',g'\>\no\\
&&=-i\int b(x)\,d^3\!x{\bigg(}[g''^{kl}(x)+g'^{kl}(x)]+i
\,b(x)^{-1}\,[\pi''^{kl}(x)-\pi'^{kl}(x)]{\bigg)}^{-1}\no\\
&&\hskip4cm\times[\delta g'^{kl}(x)-i\,b(x)^{-1}\,\delta\pi'^{kl}(x)]\no\\
&&\hskip1.8cm+i\,{\half}{\int} b(x)\,d^3\!x\,g'_{kl}(x)\,\delta g'^{kl}(x) 
\;,\en
where 
 \bn \delta g'^{kl}(x)\equiv g'^{kl}_{\;,\,j}(x)\,N^j(x)-g'^{jl}(x)\,
N^k_{\,,\,j}(x)-g'^{kj}(x)\,N^l_{\,,\,j}(x)\;, \en
and likewise for $\delta\pi'^{kl}(x)$. This relation determines the coherent 
state matrix elements of $\H_j(x)$.
Finally, observe that the diagonal coherent state matrix elements read
\bn \<\pi,g|\,\H_j(x)\,|\pi,g\>=-g_{jk}(x)\pi^{kl}_{\;\;\;|l}(x) \en
in conformity with the weak correspondence principle.

\subsection{Path integral construction}
If the given coherent states $|\pi,g\>$ possessed a resolution of unity, 
namely a nonnegative measure $\rho(\pi,g)$ (countably or even finitely 
additive) such that 
\bn \tint\<\pi'',g''|\pi,g\>\<\pi,g|\pi',g'\>\,d\rho(\pi,g)=
\<\pi'',g''|\pi',g'\>\;, \en
then the construction of a path integral for the reproducing kernel would be 
straightforward and would follow the pattern illustrated in Sec.~2 for a 
single degree of freedom. However, for the proposed reproducing kernel $
\<\pi'',g''|\pi',g'\>$ given in (39) no such measure exists and thus the 
traditional resolution of unity is unavailable. Consequently, as defined, 
$\{|\pi,g\>\}$ is a set of weak coherent states.

A similar kind of problem arose in the simple model discussed in Sec.~2 
(when $0<\beta\le\half$). In that case, the construction of a path integral 
representation proceeded in an alternative manner beginning first with a 
polarization. We assert that each of the given Hilbert space representatives,
  \bn\psi(\pi,g)\equiv\<\pi,g|\psi\>=
\sum_{n=1}^N\,\alpha_n\,\<\pi,g|\pi_n,g_n\>\in {\cal C}\;, \en
satisfies the functional differential equation [cf., (17)]
\bn&&\hskip-.5cm B^r_s(x)\,\psi(\pi,g)\equiv\bigg[-ig^{rt}(x)\frac{\delta}
{\delta\pi^{ts}(x)}+\delta^r_s+b(x)^{-1}g_{st}(x)\frac{\delta}
{\delta g_{tr}(x)}\bigg]\,\psi(\pi,g)=0 \no\\
&&\en 
for all spatial points $x$.
Next, let us introduce the operator
  \bn {\cal A}\equiv \half\int B^s_r(x)^\dagger\,B^r_s(x)\,b(x)\,d^3\!x\;, 
\en
and observe that ${\cal A}\ge0$. Thus, with $T>0$ and as $\nu\ra\infty$, 
it follows 
that ${\cal{\o J}}(\nu)e^{-\nu T {\cal A}}$, for some ${\cal{\o J}}(\nu)$, 
serves to select out the subspace where (50) is fulfilled. 
Just as in the toy model of Sec.~2, the operator ${\cal A}$ is a second order 
(functional) differential operator, and, as a consequence, a 
Feynman-Kac-Stratonovich path (i.e., functional) integral representation 
may be introduced. In particular, we obtain the formal expression [cf., (21)]
  \bn&&\hskip-1cm\<\pi'',g''|\pi',g'\>=\lim_{\nu\ra\infty}{\cal{\o N}}
\int\exp[-i\tint g_{ab}{\dot\pi}^{ab}\,d^3\!x\,dt]\no\\
 &&\hskip.3cm\times\exp\{-(1/2\nu)\tint[b(x)^{-1}g_{ab}g_{cd}{\dot\pi}^{bc}
{\dot\pi}^{da}+b(x)g^{ab}g^{cd}{\dot g}_{bc}{\dot g}_{da}]\,d^3\!x\,dt\}\no\\
&&\hskip4cm\times \prod_{x,t}\prod_{a\le b}\,d\pi^{ab}(x,t)\,dg_{ab}(x,t)\;.
\en
Here, let us interpret $t$, $0\le t\le T$,  as coordinate ``time''. On the 
right-hand side the canonical fields are functions of space and time, that is
  \bn g_{ab}=g_{ab}(x,t)\;,\hskip1cm \pi^{ab}=\pi^{ab}(x,t)\;;  \en
the overdot
($\,{\dot {~}}\,$) 
denotes a partial derivative with respect to $t$, and the integration is 
subject to the boundary conditions that $\pi(x,0),\,g(x,0)=\pi'(x),\,g'(x)$ 
and $\pi(x,T),\,g(x,T)=\pi''(x),\,g''(x)$. Observe that the field $\pi$ 
enters this path integral expression only in the form ${\dot\pi}$; this 
fact is responsible for the result of the path integral depending only on 
$\pi''-\pi'$. It is important to note, for any $\nu<\infty$, that underlying 
the formal measure given above, there is a genuine, countably additive 
measure on (generalized) functions $g_{kl}$ and $\pi^{rs}$. Loosely speaking, 
such functions have Wiener-like behavior with respect to time and 
$\delta$-correlated, generalized Poisson-like behavior with respect to space.

While (52) is invariant under spatial diffeomorphisms, it is less evident 
that it is also invariant under transformations of the time 
coordinate (by itself).\footnote{The author thanks A.~Ashtekar for raising the
question of temporal 
transformation properties.} Formally speaking, the role of the limit 
$\nu\ra\infty$ is to remove the effects of the continuous-time 
regularization. It is clear, however, that there is no need that 
removing those effects 
must be done in a {\it uniform} way independent of $x$. Thus we may replace 
$\nu$ by $\nu N(x)$---now under the integral sign---where $N(x)$, 
$0< N(x)< \infty$, is smooth and reflects the relative rate at which the 
regularization is removed at different spatial points. The end result is 
invariant under such a change. Moreover, at each point $x$ we can run the 
process with different ``clock'' rates, i.e., $N(x)\,dt\ra N(x,t)\,dt$ so 
long as the elapsed time is qualitatively unchanged. This remark means 
that we can choose any smooth {\it lapse function} $N(x,t)$, $0< N(x,t)<
\infty$, with the consequence that 
\bn&&\hskip-1cm\<\pi'',g''|\pi',g'\>=\lim_{\nu\ra\infty}{\cal N'}\int
\exp[-i\tint g_{ab}{\dot\pi}^{ab}\,d^3\!x\,dt]\no\\
 &&\hskip-.6cm\times\exp\{-(1/2\nu)\tint[b(x)^{-1}g_{ab}g_{cd}{\dot\pi}^{bc}
{\dot\pi}^{da}+b(x)g^{ab}g^{cd}{\dot g}_{bc}{\dot g}_{da}]\,N(x,t)^{-1}
\,d^3\!x\,dt\}\no\\
&&\hskip4cm\times \prod_{x,t}\prod_{a\le b}\,d\pi^{ab}(x,t)\,dg_{ab}(x,t)\;.
\en 
The necessary conditions for this more general expression to hold are, for 
all $T$, $0< T<\infty$, and at all $x$, that
   \bn && \tint_{0}^{T} N(x,t)\,dt<\infty\;, \\
   &&\tint_{0}^{\infty} N(x,t)\,dt=\infty\;. \en
In this sense we observe that our formal path integral representation (52) 
for the coherent-state overlap is actually {\it invariant} under 
transformations of the time coordinate.
\subsection{Metrical quantization}
The formal, $\nu$-dependent, weighting factor in the path integral 
expression (52) involves a {\it metric} $d\Sigma^2$ on the classical 
phase space which may be read out of the expression
  \bn  d\Sigma^2/dt^2=\tint[b^{-1}\,g_{kl}g_{rs}{\dot\pi}^{lr}
{\dot\pi}^{sk}+b\,g^{kl}g^{rs}{\dot g}_{lr}{\dot g}_{sk}]\,d^3\!x\;. \en
As presented, this expression for $d\Sigma^2$ is a {\it derived} quantity. 
Alternatively, it is clear that one could {\it start} the analysis by 
{\it postulating} a specific functional form for $d\Sigma^2$ to be used 
in a continuous-time regularization in the path integral construction of 
the reproducing kernel, and finally, by appealing to the GNS (Gel'fand, 
Naimark, Segal) Theorem \cite{gns}, to recover the representation of the 
local field operators $\sigma_{kl}(x)$ and $\kappa^r_s(x)$. Adopting a 
metric on the classical phase space as the first step in a quantization 
procedure is called {\it metrical quantization} \cite{kl7}. To carry out 
such a scheme for gravity, it is necessary that any postulated $d\Sigma^2$ 
satisfy several properties. First, it must be diffeomorphism {\it invariant} 
and second, on physical grounds, it should only depend on $d\pi^{kl}$ (or 
${\dot\pi}^{kl}$ for $d\Sigma^2/dt^2$) and not on $\pi^{kl}$ itself. Hence, 
we are initially led to consider
\bn d\Sigma^2/dt^2=\tint[b^{-2}\,L_{abcd}{\dot\pi}^{bc}{\dot\pi}^{da}+
M^{abcd}{\dot g}_{bc}{\dot g}_{da}]\,b(x)\,d^3\!x \en
for suitable, positive-definite tensors $L$ and $M$ constructed 
just from $g_{kl}$. The given choice for $L$ and $M$, i.e., $L_{abcd}=
\half[g_{ab}g_{cd}+g_{ac}g_{bd}]$ and $M^{abcd}=\half[g^{ab}g^{cd}+
g^{ac}g^{bd}]$ satisfy $M=L^{-1}$ as matrices. This choice is very natural 
and moreover is identical to the form suggested by the study of certain 
GL(3,R) coherent states for a $3\times 3$ positive-definite matrix degree
of freedom \cite{wat}. 

Nevertheless, in a metric-first quantization scheme, it is appropriate to 
examine other choices as well. For example, a term such as $b(x)^{-1}
{\dot g}_{ab}(x){\dot\pi}^{ab}(x)$ might be included, but this term may be 
eliminated by a translation of the momentum. Additionally, one may consider 
nonlocal contributions involving, for example, the term 
${\dot g}_{bc}(x){\dot g}_{da}(y)$, together with a kernel $K(x,y)$ 
specifying the interrelationship
 of the field at $y$ to the field at $x$. However, no satisfactory solution 
for $K(x,y)$ other than one proportional to $\delta(x,y)$ will lead to an 
expression for $d\Sigma^2$ that is {\it invariant} under all diffeomorphisms. 
In point of fact, the possible choices for $L$ and $M$ are rather limited, 
especially when one requires that the (formal) integration measure at each 
point is canonical and thus has the form $\Pi_{a\le b}\,d\pi^{ab}\,d g_{ab}$. 
For example, for $\lambda>0$, let
us consider  the proposal that  
\bn L_{abcd}(\lambda)\equiv\half[g_{ab}g_{cd}+g_{ac}g_{bd}+(\lambda-1)
g_{bc}g_{da}]\;.  \en
Then in order to lead to a canonical integration measure it would be 
necessary that
\bn M^{abcd}(\lambda)\equiv\half[g^{ab}g^{cd}+g^{ac}g^{bd}+
(\lambda^{-1}-1)g^{bc}g^{da}]\;.\en
Only for $\lambda=1$ is $M=L^{-1}$ which is just the choice we have made. 
(The form  for the DeWitt metric \cite{dew}, where $\lambda=-1$, is 
excluded because we require that $L$ and $M$ be positive definite.)

The preceeding discussion has rather convincingly suggested the specifically 
chosen functional form for $d\Sigma^2$---apart from one issue. It may seem 
even  more natural to choose \cite{kla77}
\bn d\Sigma^2/dt^2=\tint[g^{-1/2}g_{ab}g_{cd}{\dot\pi}^{bc}{\dot\pi}^{da}+
g^{1/2}g^{ab}g^{cd}{\dot g}_{bc}{\dot g}_{da}]\,d^3\!x \en
rather than the choice we have made. This is a natural choice from a 
classical point of view, but it is less satisfactory from a quantum point 
of view. In either case, observe that a path integral such as (52) involves 
fields with $3+1$ independent variables; however, there are no {\it space 
derivatives} involved, only {\it time derivatives}.
Such a model is known as  an {\it ultralocal quantum field theory}, and by 
now there is much that is known about the rigorous construction and 
evaluation of such nontrivial (i.e., non-Gaussian) functional integrals 
through the study, for example, of ultralocal scalar quantum fields 
\cite{kla99}. It is through the analysis of the gravitational models as 
ultralocal quantum field theories that the metric (61) is ruled out; for
a simple reason, see Section 5.

\subsection{Operator realization}
In order to realize the metric and scale fields as quantum operators in a 
Hilbert space, $\frak H$, it is expedient to introduce a set of conventional 
local {\it annihilation and creation operators},  $A(x,k)$ and  
$A(x,k)^\dag$, respectively, with the only nonvanishing commutator given by
  \bn  [A(x,k),\,A(x',k')^\dag]=\delta(x,x')\,\delta(k,k')\one\;, \en
where $\one$ denotes the unit operator.
Here, $x\in\ir^3$, while $k\equiv\{k_{rs}\}$ denotes a positive-definite, 
$3\times3$ matrix degree of freedom confined to the domain where 
$\{k_{rs}\}>0$. We introduce a ``no-particle'' state $|0\>$ such that 
$A(x,k)\,|0\>=0$ for all arguments. Additional states
are determined by suitably smeared linear combinations of
  \bn A(x_1,k_1)^\dag\,A(x_2,k_2)^\dag\,\cdots\,A(x_p,k_p)^\dag\,|0\> \en
for all $p\ge1$, and the span of all such states is $\frak H$ provided,  
apart from constant multiples, that $|0\>$ is the only state annihilated by 
all the $A$ operators. Thus we are led to a conventional Fock representation 
for the $A$ and $A^\dag$ operators. Note that the Fock operators are 
irreducible, and thus all operators acting in $\frak H$ are given as 
suitable functions of them. 

Next, let $c(x,k)$ be a possibly complex, $c$-number function and introduce 
the translated Fock operators
  \bn && \hskip.05cm B(x,k)\equiv A(x,k)+c(x,k)\,\one\;, \\
    && B(x,k)^\dag\equiv A(x.k)^\dag + c(x,k)^*\,\one\;. \en
Evidently, the only nonvanishing commutator of the $B$ and $B^\dag$ operators 
is
  \bn  [B(x,k),\,B(x',k')^\dag]=\delta(x,x')\,\delta(k,k')\one\;, \en
the same as the $A$ and $A^\dag$ operators. With regard to transformations 
of the coordinate $x$, it is clear that $c(x,k)$ (just like the local 
operators $A$ and  $B$) should transform as a scalar density of weight 
one-half. Thus we set
  \bn c(x,k)\equiv b(x)^{1/2}\,d(x,k)\;, \en
where $d(x,k)$ transforms as a scalar. The criteria for acceptable $d(x,k)$ 
are, for each $x$, that
  \bn &&\tint_+\,|d(x,k)|^2\,dk=\infty\;, \\
&&\tint_+\,k_{rs}\,|d(x,k)|^2\,dk = 2\delta_{rs}\;,  \en
the latter assuming [cf., the discussion following (39)] that 
${\tilde g}_{kl}(x)=\delta_{kl}$.
In (68) and (69) we have introduced $dk\equiv \Pi_{a\le b}\,dk_{ab}$, and 
the symbol ``$+$'' signifies an integration over only those $k$ values for 
which
$\{k_{ab}\}>0$.

We shall focus on only one particular choice for $d$, specifically,
  \bn d(x,k)\equiv \frac{ {K}\,e^{-\tr (k)}  }{\det (k)}\;, \en
which is everywhere independent of $x$; $K$ denotes a positive constant to 
be fixed later. The given choice for $d$ corresponds to the case where 
the asymptotic fields  
${\tilde\pi}^{kl}(x)\equiv0$ and 
${\tilde g}_{kl}(x)\equiv\delta_{kl}$. 
$[\!\![${\bf Remark:} For different choices of asymptotic fields it 
suffices to choose
  \bn d(x,k)\ra {\tilde d}(x,k)\equiv \frac{ {K}\,e^{ -ib(x)^{-1}
{\tilde\pi}^{ab}(x)k_{ab} }\,e^{ -{\tilde g}^{ab}(x)k_{ab} }  }
{\det (k)}\;.  \en
We shall not explicitly discuss this case further.$]\!\!]$  

In terms of these quantities, the local metric operator is defined by
 \bn \sigma_{ab}(x)\equiv b(x)^{-1}\tint_+B(x,k)^\dag\,k_{ab}\,B(x,k)
\,dk\;,\en
and the local scale operator is defined by
 \bn \kappa^r_s(x)\equiv -i\half\tint_+B(x,k)^\dag\,(k_{st}{\overrightarrow
\d^{tr}}-
{\overleftarrow\d^{rt}}k_{ts})\,B(x,k)\,dk\;.\en
Here ${\overrightarrow\d^{st}}\equiv\d/\d k_{st}$, ${\overleftarrow\d^{rt}}
\equiv\d/\d k_{rt}$ 
{\it acting to the left}, 
and $\sigma_{ab}(x)$ transforms as a 
tensor while $\kappa^r_s(x)$ transforms as a tensor density of weight one. 
It is straightforward to show that these operators satisfy the required 
affine commutation relations, and moreover that \cite{kla99,kla66}
  \bn&& \hskip.0cm\<0|\,e^{i\tint\pi^{ab}(x)\sigma_{ab}(x)\,d^3\!x}\,
e^{-i\tint\gamma^s_r(x)\kappa^r_s(x)\,d^3\!x}\,|0\> \no\\
&&\hskip.3cm=\exp\{-K^2\tint b(x)\,d^3\!x\tint[e^{-2\delta^{ab}k_{ab}}-
e^{-i\pi^{ab}(x)k_{ab}/b(x)}e^{-[(\delta^{ab}+g^{ab}(x))k_{ab}]}]\,dk/
(\det k)^2\} \no\\
&&\hskip.3cm =\exp[\!\![-2\tint b(x)\,d^3\!x\ln(\!\!(
[\det(g_{ab}(x))]^{1/2}\det\{\half[\delta^{ab}+g^{ab}(x)]-i\half b(x)^{-1}
\pi^{ab}(x)\})\!\!)\,]\!\!]\;, 
\no\\  &&\en
where $K$ has been chosen so that
  \bn K^2\tint_+ k_{rs}\,e^{-2\,\tr(k)}\,dk/(\det k)^2=2\delta_{rs}\;.  \en
An obvious  extension of this calculation leads to (39).
\subsection{Local operator products}
Basically, local products for the gravitational field operators follow the 
pattern for other ultralocal quantum field theories \cite{kla99,kla66}. 
As motivation, consider the product
  \bn  &&\hskip-1cm\sigma_{ab}(x)\sigma_{cd}(y)\no\\
&&=b(x)^{-2}\tint_+\tint_+B(x,k)^\dag\,k_{ab}\,[\,B(x,k),\,B(y,k')^\dag]
k'_{cd}\,B(y,k')\,dk\,dk'\no\\
&&\hskip2cm+:\sigma_{ab}(x)\sigma_{cd}(y):\no\\
&&=b(x)^{-2}\delta(x,y)\tint_+B(x,k)^\dag k_{ab}\,k_{cd}\,B(x,k)\,dk+
:\sigma_{ab}(x)\sigma_{cd}(y):\;, \en
where $:\;\;:$ denotes normal ordering with respect to $A$ and $A^\dag$.
When $y=x$, this relation formally becomes
\bn\sigma_{ab}(x)\sigma_{cd}(x)=b(x)^{-2}\delta(x,x)\tint_+B(x,k)^\dag 
k_{ab}\,k_{cd}\,B(x,k)\,dk+:\sigma_{ab}(x)\sigma_{cd}(x): \en
We define the renormalized (subscript ``$R$'') local product 
  \bn [\sigma_{ab}(x)\sigma_{cd}(x)]_R\equiv b(x)^{-1}\tint_+B(x,k)^\dag
\,k_{ab}\,k_{cd}\,B(x,k)\,dk \en
after formally dividing both sides by the divergent dimensionless ``scalar'' 
$b(x)^{-1}\delta(x,x)$.\footnote{For scalar ultralocal theories, the formal 
dividing factor is the divergent dimensionless ``number'' $b^{-1}\delta(0)$, 
where $b>0$ is an arbitrary factor with suitable dimensions. For gravity, 
$b\ra b(x)$, our scalar density of weight one. Note that limits involving 
test functions offer a rigorous definition of the renormalized product 
\cite{kla99,kla66}.} Higher-order local products exist as well, for example,
  \bn  &&[\sigma_{a_1b_1}(x)\sigma^{a_2b_2}(x)\sigma_{a_3b_3}(x)\cdots
\sigma_{a_pb_p}(x)]_R\no\\
&&\hskip1.5cm\equiv b(x)^{-1}\tint_+B(x,k)^\dag\,(k_{a_1b_1}k^{a_2b_2}
k_{a_3b_3}\cdots k_{a_pb_p})\,B(x,k)\,dk\,, \en
which, after contracting on $b_1$ and $b_2$, implies that  
\bn [\sigma_{a_1b}(x)\sigma^{a_2b}(x)\sigma_{a_3b_3}(x)\cdots
\sigma_{a_pb_p}(x)]_R=\delta^{a_2}_{a_1}\,[ \sigma_{a_3b_3}(x)\cdots
\sigma_{a_pb_p}(x)]_R \;.\en
It is in this sense that $[\sigma_{ab}(x)\sigma^{bc}(x)]_R=\delta^c_a$. 

We take up only one further point regarding local products. It is rather 
natural \cite{kla99,kla66}
to try to define the local momentum ``operator'' by
  \bn \mu^{rs}(y)=-i\half\tint_+ B(y,k)^\dag\,({\overrightarrow\d^{ab}}
- -{\overleftarrow\d^{ab}})
\,B(y,k)\,dk\;, \en
but this expression only leads to a form and not a local operator. 
Furthermore, the putative canonical commutation relation becomes
  \bn &&[\sigma_{ab}(x),\,\mu^{rs}(y)]=i\,\delta^{rs}_{ab}\,\delta(x,y)
\,b(x)^{-1}\,\tint_+B(x,k)\,B(x,k)\,dk\no\\
  &&\hskip2.8cm=i\,\delta^{rs}_{ab}\,\delta(x,y)\,[\tint_+|d(x,k)|^2\,dk+
\ldots]\;,\en
which has a divergent multiplier and is, therefore, not even a form. On the 
other hand, it is true that
  \bn &&\half[\sigma_{rl}(x)\mu^{ls}(x)+\mu^{sl}(x)\sigma_{lr}(x)]_R\no\\
&&\hskip1cm=-i\half\tint_+B(x,k)\,(k_{rl}{\overrightarrow\d^{ls}}
- -{\overleftarrow\d^{sl}}k_{lr})\,B(x,k)
\,dk\no\\
&&\hskip1cm =\kappa^s_r(x) \en
as claimed.
\section{Imposition of Constraints}
Gravity has four constraints at every point $x\in{\cal S}$, and, when 
expressed in suitable units, they are the familiar spatial and temporal 
constraints, all densities of weight one, given by \cite{mtw}
 \bn && H_a(x)=-g_{ab}(x)\pi^{bc}_{\;\;\;|c}(x)\;,  \\
&&H(x)=\half g(x)^{-1/2}[g_{ab}(x)g_{cd}(x)+g_{ad}(x)g_{cb}(x)-2g_{ac}(x)
g_{bd}(x)]\no\\ &&\hskip2.5cm\times\pi^{ac}(x)\pi^{bd}(x)+g(x)^{1/2\;
\;(3)}\!R(x)\;. \en
The spatial constraints are comparatively easy to incorporate since their 
generators serve as generators of the diffeomorphism group acting on 
functions of the canonical variables. Stated otherwise, finite spatial
diffeomorphism 
transformations  map any coherent state onto another coherent state as in 
(41) and (42). However, this is decidely not the case for the temporal 
constraint. What follows is an account of what to do about these constraints 
{\it in principle}; in Part II on this subject, we will discuss how to 
accomplish these goals.

One satisfactory procedure to incorporate all the necessary constraints 
is as follows. Let $\{h_p(x)\}_{p=1}^\infty$ denote a complete, orthonormal 
set of real functions on $\cal S$ relative to the weight $b(x)$. In 
particular, we suppose that
  \bn &&\tint h_p(x)\,h_n(x)\,b(x)\,d^3\!x=\delta_{pn}\;,  \\
  &&\hskip0.05cm b(x)\,\Sigma_{p=1}^\infty\,h_p(x)\,h_p(y)=\delta(x,y)\;.\en
Based on this orthonormal set of functions, we next introduce four infinite 
sequences of constraints
  \bn  &&\hskip-.2cm H_{(p)\,a}\equiv\tint h_p(x)\,H_a(x)\,d^3\!x\;, \\
   &&H_{(p)}\equiv\tint h_p(x)\,H(x)\,d^3\!x\;,  \en
$1\le p<\infty$, all of which vanish in the classical theory. 

For the quantum theory let us assume, for each $p$, that $\H_{(p)\,a}$ and 
$\H_{(p)}$ are self adjoint, and even stronger that 
  \bn X_P^2\equiv \Sigma_{p=1}^P \,2^{-p}\,[\Sigma_{a=1}^3(\H_{(p)\,a})^2+
(\H_{(p)})^2] \en
is  self adjoint for all $P<\infty$. Note well, as one potential example, 
the factor $2^{-p}$  introduced as part of a regulator as $P\ra\infty$; we 
comment on this regulator in the next section. For each $\delta\equiv
\delta(\hbar)>0$, let
  \bn \E_P\equiv \E(\!\!(X_P^2\le\delta^2)\!\!) \en
denote a projection operator depending on $X_P$ and $\delta$ as indicated. 
How such projection operators may be constructed is discussed in \cite{kla2} 
and will be dealt with in Part II. 
Let 
   \bn S_P\equiv \limsup_{\pi,\,g}\,\<\pi,g|\E_P|\pi,g\>\;, \en
which satisfies $S_P>0$ since $\E_P\not\equiv0$ when restricted to 
sufficiently large $\delta$. 
 Finally, we  define
   \bn\<\!\<\pi'',g''|\pi',g'\>\!\>\equiv \limsup_{P\ra\infty}\,S_P^{-1}\,
\<\pi'',g''|\E_P|\pi',g'\>  \en
as a reduction of the original reproducing kernel.
The result is either trivial, say if $\delta$ is too small, or it leads to 
a continuous, positive-definite functional on the original phase space 
variables. We focus on the latter case.

To obtain the final physical Hilbert space, one must study 
$\<\!\<\pi'',g''|\pi',g'\>\!\>$ as a function of the regularization parameter 
$\delta$. Since gravity has an anomaly \cite{and}, there should be a 
minimum value of 
$\delta$, which is still positive, that defines the proper theory, rather 
like the example in (30). Assuming we can find and then use that value, 
$\<\!\<\pi'',g''|\pi',g'\>\!\>$ becomes the reproducing kernel for the 
physical Hilbert space ${\frak H}_{\rm\,phys}$. Attaining this goal would 
then permit the real work of extracting the physics to begin.

Our discussion regarding constraints in this paper has indeed been brief. 
Although the program we have in mind is not simple, it has the virtue of 
being realizable, at least in principle. After all, before any calculational 
scheme is developed it is always wise to ensure that the object under study 
has a good chance of existing!

\section{Discussion}
In the preceding sections we have outlined an approach to quantum gravity 
that it is somewhat different than currently considered. As background to our 
philosophy, let us briefly review some of the common weak points in the 
standard ways of quantizing gravity, and use these comments as motivation 
for our approach.
\subsection{Traditional viewpoints \& commentary}

(i) Viewed by way of conventional perturbation theory, quantum gravity has 
two main difficulties of principle. On the one hand, the perturbative split 
of the metric in the form $g_{\mu\nu}(x)=\eta_{\mu\nu}+h_{\mu\nu}(x)$ (or 
any other background metric), with canonical quantization of the ``small'' 
deviation $h_{\mu\nu}(x)$ violates signature properties since in that case 
the spectrum of $h_{\mu\nu}(x)$ is unbounded above and below. On the other 
hand, as an asymptotically nonfree theory, gravity is nonrenormalizable and 
poorly described by a perturbation theory which needs an unending addition of 
distinct counterterms with divergent coefficients. 

To address these obstacles, we first note that the affine approach guarentees 
a proper metric signature from the very beginning, and second we remind the 
reader that certain asymptotically nonfree, nonrenormalizable models have 
indeed been solved \cite{kla99}, and their solution procedures form the core 
of the present approach to quantize gravity.\vskip.2cm

\noindent(ii) While the constraints of classical gravity are first class, 
there is an anomaly in the quantum constraints and thus they are effectively 
second class. Usual views toward second-class constraints involve solving and 
eliminating them, introducing and then quantizing Dirac brackets, or the 
conversion of second-class constraints into 
first-class constraints. Each of these methods is often complicated and 
not all are guaranteed to be valid beyond a semiclassical treatment if the 
classical constraint hypersurface has a non-Euclidean geometry \cite{shabig}. 
These difficulties have stimulated searches to get around the second-class 
character altogether, either by introducing non-Hermitian constraint 
operators that may close algebraically \cite{kom}, or by introducing 
additional fields 
and space-time dimensions until the anomaly cancels.

Regarding these comments, we accept the anomaly and the second-class 
constraints that it implies. Giving up a classical symmetry is is not so 
heretical as it may seem. For example, Hamiltonian classical mechanics 
enjoys a full covariance under general canonical coordinate transformations, 
but that invariance is {\it not} preserved {\it in its classical form} 
when we go to the quantum theory. For example, consider the classical 
Poisson brackets for a set of generator elements
  \bn \{e^{ap+bq},\,e^{cp+dq}\}=(bc-ad)\,e^{(a+c)p+(b+d)q}\;, \en
where $a,b,c$, and $d$ are parameters, 
while in canonical quantum mechanics we have the corresponding 
commutator algebra
\bn [e^{aP+bQ},\,e^{cP+dQ}]=(2i)\sin[(bc-ad)/2]\,e^{(a+c)P+(b+d)Q}\;. \en
These expressions agree in their algebraic structure for selected elements, 
but not for the whole algebra. Equivalence begins to break down at the 
quadratic level, which is exactly the case for the temporal constraint 
in gravity. In particle mechanics there are sound physical reasons 
\cite{kla87} for this ``breakdown'' of symmetry, and attempts to restore 
the symmetry---as in geometric quantization \cite{geo}---go counter to 
such sound physical principles. There is no reason that a similar scenario 
does not hold for gravity. The breakdown of the classical symmetry and the 
appearance of a quantum ``anomaly'' (better called a ``quantum mechanical 
symmetry breaking'' \cite{jac}) could, just as in the quantum mechanics case, 
carry real physics. 

Accepting the second-class nature of (part of) the constraints of 
quantum gravity means a different 
approach must be taken. As already noted, earlier approaches required 
solving for and 
eliminating the unphysical variables, the introduction of Dirac brackets, 
etc., all of which are rather technical and may be extremely complicated. 
In the present view, afforded by the projection operator approach to 
constraints---{\it second-class constraints included}---none of these 
particular complications arise. Instead, one projects onto the state (or, 
with degeneracy, states) for which the sum of the square of the constraints 
is bounded. Why the square and not the fourth power? Using the fourth power 
would not be wrong; the only change would involve a unitary transformation 
of the original result, which maps one set of ``ground'' states onto another 
set of ``ground'' states. The square is chosen for simplicity, not for any 
reasons of exclusivity. 

\subsection{Relation to earlier work}
Pilati, in a series of papers \cite{pli} (see also \cite{ish}), analyzed a 
strong coupling model of quantum gravity in which the temporal constraint 
given in (85) was modified to read
  \bn &&\hskip-.8cm H'(x)=\half g(x)^{-1/2}[g_{ab}(x)g_{cd}(x)+g_{ad}(x)
g_{cb}(x)-2g_{ac}(x)
g_{bd}(x)]\,\pi^{ac}(x)\pi^{bd}(x)\;,\no\\
  && \en
namely, the second term involving the scalar curvature $\!{~}^{(3)}\!R(x)$ 
based on the metric $g_{ab}(x)$ was dropped. The reason for doing so was to 
achieve a theory in which the temporal constraint $H'(x)$ itself  was 
patterned after the  Hamiltonian density of an ultralocal theory. This 
modification  was thought to be advantageous because then all the machinery 
developed for ultralocal quantum field theory could be used for the strong 
coupling gravitational model. Once that model was under control, it was the 
hope to reintroduce the dropped term by a perturbation theory analysis. 
Unfortunately, the reintroduction of dropped terms involving spatial 
gradients has never been successfully accomplished by a perturbation 
analysis about a non-Gaussian ultralocal model. This failure is most 
likely because such ``interaction terms'' generally amount to 
nonrenormalizable perturbations of the unperturbed (ultralocal) models.

The program advanced in the present paper takes a different view toward 
these issues.

First, we focus on kinematics with the knowledge that for pure gravity 
the Hamiltonian operator vanishes, as it does in any situation that is 
reparametrization invariant. In its place we find constraints, and the 
real physical content of the theory lies in the particular constraints. 
However, before the constraints can be introduced, there must be a 
``primary container'' to receive them. In our case, this primary container 
is the Hilbert space and set of relevant operators prior to the introduction 
of {\it any} form of the constraints, and which is based on the fundamental 
physical nature of the variables, i.e., positive-definite,  $3\times3$ 
matrix-valued, local field operators,  etc. This is the preferred procedure: 
{\it Quantization before the introduction of any constraints}. At this 
primary level, there is no coupling of one degree of freedom with 
another---any coupling comes through the enforcement of specific constraints. 
Hence, in the primary container the degrees of freedom are mutually 
independent of each other. For finitely many kinematical degrees of 
freedom this means that the Hilbert space is a product over spaces for 
each of the separate degrees of freedom; in a field theory, this 
independence means that the kinematical operators enter as ultralocal 
field operators. Consequently, even though the several constraint operators 
waiting to be introduced may themselves {\it not} be ultralocal in nature, 
the primary container itself, which has been prepared to receive them, is 
ultralocal.

At this point the reader may wish to reexamine (39)---in essence our 
``primary container''---to recall the appearance of an ultralocal state 
on a set of field operators. Apart from the $3\times3$ matrix character, 
the functional form of (39) emerges from (i) the product of $N$ expressions 
of the form (16) for independent arguments $p_n,q_n,r_n,s_n$, $1\le n\le N$, 
(ii) the replacement of $\beta$ by $b_n\Delta$ and $(p_n-r_n)$ by $(p_n-r_n)
\Delta$, and (iii) the limit as $N\ra\infty$, $\Delta\ra0$ such that 
$\Sigma (\cdot)\,b_n\,\Delta\ra\tint (\cdot)\,b(x)\,dx$. In this way we 
have created, from a collection of independent single affine degrees of 
freedom, the reproducing kernel for affine gravity in $1+1$ dimensional 
space. In a similar manner, a set of independent $3\times3$ affine degrees 
of freedom can be (and were) used to build an ultralocal representation for 
$3\times3$ metric and momentum fields in (39); moreover, this type of 
construction does not favor the ``natural'' phase-space metric (61). 
In summary, we emphasize that whenever the ``dynamics'' appears through 
constraints, the primary container should be ultralocal in character. We 
next turn our attention to the introduction of the constraints.

In the projection operator approach it is recognized from the outset that 
the physical Hilbert space---or better the {\it regularized} physical 
Hilbert space---is a {\it subspace} of the original Hilbert space that is 
uniquely determined by an associated projection operator $\E$. Whatever 
form the constraints may take, they are ``encoded'' into the projection 
operator $\E$, and a regularization means that the constraints are 
satisfied to a certain level of precision determined by  a regularization 
parameter $\delta$. How to turn constraint operators into projection 
operators in general has been discussed in \cite{kla2}; as regards the 
gravitational case, that project will be discussed in Part II .  

Continuing still in a general framework, let us consider an expression that 
may be used either to generate dynamics or to enforce constraints. From a 
classical point of view, and especially from a path integral point of view, 
it may seem that  quantities used in either of these ways may be rather 
similar. However, it is important to already understand that there is a 
fundamental distinction between the use of a quantum operator either (i) 
to generate unitary transformations or (ii) to serve as a constraint operator 
in a
given system.  In the first case, the operator must be self adjoint and 
thus densely defined, while in the second case the operator may be defined 
on only the zero vector! This fact has profound consequences. In particular, 
to have a self-adjoint generator requires that the operator representation 
in the primary container must already be finely ``tuned'' to ensure that the 
generator which will be introduced is self adjoint (as in Haag's Theorem 
\cite{hag}). For constraint operator imposition this need not be the case, 
and the reason this is so is because we allow for changes, i.e., adaptions, 
of the primary container representations through the process of reduction 
of the reproducing kernel. As an example, let us consider only the local 
temporal operator $\H(x)$ for gravity. On the one hand, to generate unitary 
time evolutions it may be necessary that $\tint\H(x)\,d^3\!x$ be self 
adjoint. On the other hand,  to enforce constraints, it is only necessary 
that $\H_{(p)}$ [cf., (89)] be ``small'', but there is no requirement that 
these operators must be {\it uniformly} ``small''. Instead they can be 
``small'' in the sense that $X^2\equiv\Sigma_{p=1}^\infty s_p (\H_{(p)})^2$ 
is ``small'', where the set of positive constants $\{s_p\}$ serve as 
regulators to control convergence of the series. The example chosen was 
that $s_p=2^{-p}$, but there is nothing special about that choice. Any 
reasonable choice that leads to a self adjoint operator $X$ should lead 
to the same reproducing kernel in the final analysis when the regularization 
parameter $\delta$ attains its final value for the problem at hand.

A rather simple example of the general procedure discussed above can be seen 
in studies involving product representations \cite{fad}. Additionally, it 
is instructive to reanalyze the relativistic free field by this procedure 
to see how ultralocal representations turn into nonultralocal representations.

Suffice it to say, it is this vast difference between the required nature of 
constraint operators and unitary generators that permits us to start with 
ultralocal field operator representations and emerge with {\it non}ultralocal 
operators in the physical Hilbert space. 
\section*{Acknowledgments}
The author thanks B.~Bodmann, J.~Govaerts, A.~Kempf, S.V.~Shabanov, and 
G.~Watson for useful comments.

\end{document}